     \documentstyle[12pt]{article}

     \newcommand{\eq}{\begin{equation}}

     \newcommand{\en}{\end{equation}}

     \newcommand{\spa}{\hspace{0.5cm}}
     \newfont{\myfont}{msbm10 scaled\magstep1}
     \newcommand{\del}{\partial}
     \newcommand{\csi}{\epsilon^{\mu\alpha\nu}{\cal A}_\mu
\partial_\alpha
{\cal A}_\nu}

     \newcommand{\cs}{\epsilon^{\mu\alpha\nu}A_\mu
\partial_\alpha
A_\nu}
      \newcommand{\cmu}{\epsilon^{\mu\alpha\nu}\partial_\alpha}

    \newcommand{\expo}{\exp \{ - \int d^3z [}

  \newcommand{\intd}{\int d^3z_E\  d^3z'_E}

   \newcommand{\ab}{\bar{A}}

\newcommand{\fb}{\bar{F}}

\newcommand{\ca}{{\cal A}}

\newcommand{\cf}{{\cal F}}

\newcommand{\cl}{{\cal L}}

\newcommand{\fr}{\frac}

\begin{document}

      \begin{titlepage}

\rightline{PUPT-1368}
\rightline{December 1992}
\vskip 10.0mm

 \centerline{ \LARGE\bf Quantum Electrodynamics of Particles}
 \vskip 3pt

 \centerline{\LARGE\bf on a Plane and the Chern-Simons Theory}

    \vskip 2.0cm

    \centerline{\sc E.C.Marino \footnote{On sabbatical leave from
Departamento de F\'{\i}sica, Pontif\'{\i}cia Universidade
Cat\'{o}lica, Rio de Janeiro, Brazil. E-mail :
marino@puhep1.princeton.edu }}

     \vskip 0.6cm

     \centerline{\it Joseph Henry Laboratories }
    \centerline{\it   Princeton University }
    \centerline{\it    Princeton, NJ 08544 }

     \vskip 1.8cm

  \begin{abstract}

 We study the electrodynamics of generic charged particles
(bosons, fermions, relativistic or not) constrained to move
on an infinite plane. An effective gauge theory
in 2+1 dimensional spacetime which describes the real electromagnetic
interaction of this particles is obtained. The relationship between
this effective theory with the Chern-Simons theory is explored. It
is shown that the QED lagrangian {\it per se} produces the
Chern-Simons constraint relating the current to the effective
gauge field in 2+1 D. It is also shown that the geometry of the
system unavoidably induces a contribution from the topological
$\theta$-term that generates an explicit Chern-Simons term
for the effective 2+1 dimensional gauge field as well as a minimal
coupling of the matter to it. The possible relation of the effective
three dimensional theory with the bosonization of the Dirac fermion
field in 2+1 D is briefly discussed as well as the potential
applications in Condensed Matter systems.

  \end{abstract}
\vfill
\end{titlepage}

\leftmargin 25mm

\topmargin -8mm
\hsize 153mm

\baselineskip 7mm
\setcounter{page}{2}

\leftline{\Large\bf 1) Introduction}
\bigskip

\spa In the last few years there has been an intense theoretical
activity focused in the investigation of the  properties of
 charged particles constrained to move
on a plane. The main reason for this certainly comes from
the recent discoveries of the Quantum Hall Effect \cite{qhe} and
of the High Temperature Superconductivity \cite{sc} in the
eighties. In both cases the fundamental physical properties
of the system seem to be completely determined by the dynamical
behavior of a two dimensional array of electrons.

\spa In all of the interesting realistic physical
applications, in spite of the
fact that the motion of the electrons is confined to a plane,
of course the electromagnetic fields through which they
interact are certainly not subject to this constraint. The electrostatic
potential between two electrons on a plane,
 for instance, is the familiar
1/r Coulomb potential despite their planar motion and not the
logarithmic potential that would be peculiar of electrodynamics
in 2+1 dimensions.

\spa Nevertheless, the obtainment from first
principles in 3+1 D, of a complete
 2+1 dimensional description of this kind of electronic
systems which move on a plane
but interact, of course, as the
four-dimensional particles which they are, would be
highly convenient.
One would
immediately wonder, in particular, what would be the
relationship between this effective description and the strict
2+1 dimensional theories that have been constructed in order to explain
these systems.

\spa There is a field theory which plays a central role in the
set of models built to describe the properties of condensed matter
systems in 2+1 D. This is the Chern-Simons theory \cite{cs}. In the case
of the Hall Effect, it leads very naturally to the
transverse conductivity characteristic of this effect. Another
important property of the Chern-Simons theory is that it
induces a change in the statistics of the particles coupled to it,
the so called statistical transmutation \cite{transmut}.  A system of
generalized statistics particles or anyons, on the other
hand was shown to spontaneously have a superconducting
ground state, even in the absence of the attractive
phonon interactions which lead to the usual BCS superconductivity
\cite{sup}. Hence the potential interest in Chern-Simons
theories in connection with  High-Tc Superconductivity.

\spa The purpose of this work is to obtain an effective
2+1 dimensional field theory describing the real (3+1 D)
electromagnetic interaction of generic charged particles
(fermionic, bosonic, relativistic or not) constrained to
move on an infinite plane. We show that the Maxwell
lagrangian of QED leads {\it per se} to an effective 2+1 D
theory which possesses the Chern-Simons constraint, being
therefore intimately related to this theory. We also show
that the geometry implied by the restriction of the motion
to a plane unavoidably induces a contribution from the
topological $\theta$-term \cite{bau} which will generate the
minimal coupling of
matter to the effective gauge field in 2+1 D (``statistical field'')
responsible for the statistical transmutation  as well as an
explicit Chern-Simons term.

\spa It is important to notice the fundamental difference
between the Chern-Simons term which we obtain here which
has its origins strictly in the 3+1 dimensional gauge fields
and the one which is obtained by integration over the
matter fields \cite{csgeneration}. The latter, for instance
is no longer able to produce a statistical transmutation
because after the integration over the matter fields there
no longer particles interacting with the Chern-Simons field.

\spa The work is organized as follows. In section 2 we review
the main properties of the Chern-Simons theory. In section 3
we obtain the effective electromagnetic interaction for particles
constrained to move on a plane. In section 4 we show how to
 produce this interaction  within a 2+1 D field theory. In
section 5 we show that the effective gauge field theory in
2+1 D presents the same constraint as the Chern-Simons theory.
In section 6 we show how the geometry of the system
induces a contribution from the $\theta$-term of QED
which will lead to a Chern-Simons explicit term.
The conclusions are presented in section 7.

\bigskip
\leftline{\Large\bf 2) The Chern-Simons Theory}
\bigskip

\spa The Chern-Simons theory is characterized by the following lagrangian
in 2+1 dimensions
$$
{\cal L}=\frac{\theta}{2}\csi -q\  j^\mu {\cal A}_\mu +{\cal L}_M+
{\cal L}_{GF}
\eqno (2.1)
$$
where ${\cal L}_M$ is the matter lagrangian and ${\cal L}_{GF}$ is a
gauge fixing term. Also in the above equation, $j^\mu$ is the
matter current in 2+1 D (which has dimension 2) and $q$ is the
(dimensionless) coupling of matter to the gauge field ${\cal A}_\mu$
which is usually called  the statistical field. $\theta$ in (2.1)
is an arbitrary real parameter, the ``statistical parameter''.
The field equation associated to (2.1), namely
$$
j^\mu = \frac{\theta}{q} \cmu {\cal A}_\nu
\eqno (2.2)
$$
implies that the statistical field is completely determined
by the matter current. The field equation is actually a constraint.
In the Lorentz gauge, we obtain from (2.2) the expression
for $\ca_\mu$ in terms of the current
$$
{\cal A}^\mu = \frac{q}{\theta}\cmu [\frac{1}{-\Box}] j_\nu
\eqno(2.3)
$$
The constraint (2.2) implies the following relation between the
matter current and the electric and magnetic statistical fields
$$
j^0 =\frac{\theta}{q}{\cal B}
\eqno(2.4a)
$$
$$
j^i =\frac{\theta}{q}\epsilon^{ij}{\cal E}_j
\eqno (2.4b)
$$
The first of this equations
leads to the well known statistical transmutation of the matter fields,
the change in statistics being given by $\Delta s=\frac{q^2}{2\pi\theta}$
\cite{transmut}. This happens because (2.4a) implies that a
magnetic flux $q/\theta$ is attached to each particle having a charge
$q$ \cite{wil}.
 The second one, on the other hand,
implies a transverse conductivity which underlies the usefulness of
Chern-Simons theories in the description of the Hall effect. We see
that the Hall conductance would be given by $\theta$ (of course,
in any realistic theory of the Hall effect, we should have the
real electric field in (2.4b)!).

\spa The Chern-Simons theory produces a peculiar effective long
range interaction between matter particles which is, after all, the
responsible for the statistical transmutation. We can get this
effective interaction by considering the result
of the integration over $\ca_\mu$ (in euclidean space)
$$
Z^{CS}_{eff}=Z^{-1} \int D{\cal A}_\mu \expo -\frac{i\theta}{2} \csi
+q\ j^\mu {\cal A}_\mu +{\cal L}_M
$$
$$
-\frac{\xi\theta}{2}{\cal A}^\mu \frac{\del^\mu\del^\nu}{(-\Box)^{1/2}}
{\cal A}_\nu ] \}
 \eqno (2.5)
$$
The last term in the exponent above is the properly chosen gauge fixing
term. Multiplication by  $(-\Box)^{-1}$ of course is to be
understood in the convolution sense. Integrating over the statistical
field with the help of the euclidean propagator of this field,
namely
$$
G^{\mu\nu} = \frac{i}{\theta}\cmu [\frac{1}{-\Box_E}]
-\frac{1}{\xi\theta} \del^\mu\del^\nu[\frac{1}{(-\Box_E)^{3/2}}]
\eqno(2.6)
$$
we get
$$
Z^{CS}_{eff}=\exp \{\frac{q^2}{2}\intd\  j^\mu(z) G_{\mu\nu}(z-z')j^\nu(z')-
\int d^3z {\cal L}_M \}
$$
$$
=\exp \{ -\int d^3z_E [ {\cal L}_M-i\frac{q^2}{2\theta}
 \epsilon^{\mu\alpha\nu}j_\mu\del_\alpha[\frac{1}{-\Box_E}]j_\nu] \}
\eqno(2.7)
$$
The last term in
the exponent of the above expression is the ``statistical interaction'',
the long range interaction which is responsible for the change in statistics
 of the matter particles.

\bigskip
\leftline{\Large\bf 3)The Effective Electromagnetic Interaction}
 \leftline{\Large\bf\hskip 5.0mm for Particles on a Plane}
\bigskip

\spa Let us consider Quantum Electrodynamics in 3+1 D for generic
charged particles
$$
{\cal L}_{EM} =-\frac{1}{4}F^2_{\mu\nu} -e j^\mu_{3+1}A_\mu
+{\cal L}_M +{\cal L}_{GF}
\eqno (3.1)
$$
In the above expression, ${\cal L}_M$ is the matter kinetic lagrangian which
is completely arbitrary: it may be relativistic or not and the particles
it
describes may be either fermions or bosons.
$\cl_{GF}$ is the gauge fixing term, $j^\mu_{3+1}$ is the matter
current in 3+1 D and $e$ is the charge of the matter particles.

\spa The electromagnetic field induces an effective interaction of
matter which can be obtained by integrating over $A_\mu$ in (3.1).
Going to euclidean space and using the euclidean propagator of
the electromagnetic field, namely
$$
G_{EM}^{\mu\nu} = [-\Box_E \delta^{\mu\nu}+(1-\frac{1}{\xi})\del^\mu\del^\nu]
[\frac{1}{(-\Box_E)^2}]
\eqno (3.2)
$$
we get
$$
Z^{EM}_{eff} = Z^{-1} \int DA_\mu \exp \{- \int d^4z_E [\fr{1}{4}
 F^2_{\mu\nu} + e\ j^\mu_{3+1}A_\mu -\fr{\xi}{2}A_\mu \del^\mu\del^\nu
A_\nu ]\}
$$
$$
= \exp \{ \fr{e^2}{2} \int d^4z_E\ d^4z'_E\ j^\mu_{3+1}(z)\ G^{\mu\nu}
_{EM}(z-z')\ j^\nu_{3+1}(z') \}
$$
$$
= \exp \{\frac{e^2}{2}\int d^4z_E\ d^4z'_E\  j^\mu_{3+1}(z)[\frac{1}{-\Box_E}]
j^\mu_{3+1}(z')\}\equiv \exp[-S_{eff}[j^\mu_{3+1}]]
 \eqno (3.3)
$$
Observe that only the first term of (3.2) contributes to (3.3) due
to current conservation. The effective action in (3.3) is the
familiar electromagnetic interaction. For static point charges,
for instance, the energy corresponding to it is the familiar
1/r Coulomb potential
energy.

\spa Since we are interested in describing particles in 2+1 D, i.e.,
matter confined to a plane, let us consider the case in which the
current is given by
$$
j^\mu_{3+1}(x^0,x^1,x^2,x^3)=\cases{j^\mu(x^0,x^1,x^2)\delta(x^3)&$
\mu =0,1,2$\cr 0 &$ \mu =3 $\cr }
\eqno (3.4)
$$
Inserting this expression (euclideanized) in the effective action in (3.3) and
integrating over $z^3 $ and $z'^3$ we get
$$
S_{eff} =-\frac{e^2}{2}\intd\  j^\mu(z) K_E(z-z'|z^3=z'^3=0)j^\mu(z')
\eqno (3.5)
$$
where the euclidean kernel $K_E$ is given by
$$
K_E(z-z'|z^3=z'^3=0)\equiv [\frac{1}{(-\Box)_E}]_{z^3=z'^3}
$$
$$=
\int \frac{d^4k}{(2\pi)^4}\frac{\exp [i \sum_{i=1}^4 k\cdot (z_i-z_i')]}
{\sum_{i=1}^4\ k_i^2}\ |_{z_3=z_3'=0}
=\frac{1}{8\pi^2 \sum_{i=1,2,4} (z_i-z'_i)^2}
\equiv \frac{1}{8\pi^2 |z-z'|^2_{3D}}
\eqno(3.6)
$$
where we have defined the 3-D vectors $z_{3D}=(z_1,z_2,z_4)$.

\spa Equations (3.5) and (3.6) determine the electromagnetic interaction
of matter which is confined to a plane as is implied by the
expression for the  current, eq.(3.4).
 We shall see in the next section how this effective interaction
can be obtained from a theory in 2+1 dimensional spacetime.

\vfill
\eject

\leftline{\Large\bf 4) The Pseudo Electromagnetic Field}
\bigskip

\spa Let us start by observing that expression (3.6) for the effective
kernel can be written as a three-dimensional integral
$$
\frac{1}{8\pi^2|z-z'|^2_{3D}} = \frac{1}{4} \int \frac{d^3k_{3D}}{(2\pi)^3}
\frac{e^{i [k\cdot (z-z')]
_{3D}}}{(k_{3D}^2)^{1/2}} \equiv\fr{1}{4}\  \fr{1}{(-\Box_{2+1}^E)^{1/2}}
\eqno(4.1a)
$$
where $\Box_{2+1}^E $ is the (euclidean) three-dimensional d'Alembertian
operator.
Replacing (3.6) for (4.1a) and inserting in
(3.5) we can see that the effective electromagnetic
interaction for the charged particles on a plane is already
completely expressed within a three-dimensional world:
$$
S_{eff} =-\frac{e^2}{8}\intd\  j^\mu(z)[\fr{1}{
(-\Box^E_{2+1})^{1/2}}](z-z')j^\mu(z')
\eqno(4.1b)
$$

\spa We will show in this section that we can
 obtain this effective electromagnetic interaction by starting
from a three-dimensional field theory in which the 2+1 D
matter current $j^\mu$ interacts with the 2+1 D gauge field $\ab_\mu$,
which we call the pseudo
electromagnetic field. Let us consider the following lagrangian
for the $j^\mu-\ab_\mu$ system in 2+1 dimensions
(we henceforth drop the subscript ``2+1'' in
the three dimensional d'Alembertian)
$$
{\cal L}_{PEM}
=-\frac{1}{4}\fb_{\mu\nu}[\frac{1}{\Box^{1/2}}]\fb^{\mu\nu}
-\frac{e}{2}j^\mu\ab_\mu +\frac{\xi}{2}\ab_\mu\frac{\del^\mu\del^\nu}
{\Box^{1/2}}\ab_\nu
\eqno(4.2)
$$
where the last term above is a gauge fixing term.

\spa Going to euclidean space and integrating over the $\ab_\mu$
field, with the help of its euclidean propagator, namely
$$
D^{\mu\nu} = [-\Box_E \delta^{\mu\nu} +(1-\frac{1}{\xi})\del^\mu\del^\nu]
[\frac{1}{(-\Box_E)^{3/2}}]
\eqno(4.3)
$$
we get
$$
Z_{eff}^{PEM}\equiv \exp[-S_{eff}] =\exp \{ \frac{e^2}{8} \intd\
j^\mu(z) [\frac{1}{(-\Box_E)^{1/2}}] j^\mu(z') \}
\eqno(4.4)
$$
Observe that because of current conservation, only the first
term of (4.3) contributes in (4.4). Comparing with (4.1b),
we conclude, therefore, that the theory described by (4.2)(in 2+1 D)
reproduces precisely the same  effective interaction which was
obtained from QED after imposing that the charges are confined to a
plane.
This fact was shown here in euclidean space. Going to Minkowski
space, one can show \cite{rub} that the identity (4.1a) is also valid.
Choosing a certain prescription (e.g. Feynman, advanced
, etc) for the kernel in the
$\ab_\mu$-field theory would lead to the corresponding kernels
of QED. In particular, it is interesting to note that the
theory described by (4.2) in 2+1 D would produce a Coulombic
(1/r) interaction between static charges in the plane, which is
the correct result for real charges, instead of the logarithmic(unphysical)
potential which is  known to be produced by three-dimensional QED.

\spa  The theory described by (4.2) was proposed to be associated with
 the bosonization of the free Dirac fermion field in 2+1 D \cite{bos}.
As we shall comment later, it is rather suggestive that it also appears
in the present context.

\bigskip
\leftline{\Large\bf 5) The Statistical Field and the Chern-Simons}
\vskip 3pt
\leftline{\Large\bf\hskip 5.0mm Constraint}
\bigskip

\spa Let us show here that the theory of the pseudo electromagnetic field
defined by (4.2) is equivalent to the following theory involving
the gauge field $\ca_\mu$ which for reasons that will become clear
later, will shall call the statistical field. Consider
$$
\cl [\ca_\mu,\ab_\mu] = -\fr{\lambda^2}{4}\cf_{\mu\nu}
[\fr{1}{\Box^{1/2}}]
\cf^{\mu\nu} +\lambda \epsilon^{\mu\alpha\beta}\ab_\mu \del_\alpha
\ca_\beta + \lambda^2
\frac{\xi}{2}\ca_\mu\frac{\del^\mu\del^\nu}
{\Box^{1/2}}\ca_\nu
\eqno(5.1)
$$
where the last term is the gauge fixing and $\lambda$ is an arbitrary
real constant.

\spa Going to euclidean space and integrating over the $\ca_\mu$
field we get the effective $\ab_\mu$ action:
$$
Z_{eff}[\ab_\mu] \equiv \exp \{-S_{eff}[\ab_\mu]\}= Z_0^{-1}
\int D\ca_\mu \expo \fr{\lambda^2}{2}\ca_\mu
[\fr{-\Box_E\delta^{\mu\nu}+(1-\xi)\del^\mu\del^\nu}{(-\Box_E)^{1/2}}]
\ca_\nu
$$
$$
-i\lambda \epsilon^{\mu\nu\alpha}\ab_\mu\del_\nu\ca_\alpha] \}
$$
$$
= \exp \{ -\fr{\lambda^2}{2}\intd\  \ab_\mu(z) \epsilon^{\mu\sigma
\alpha}\del_\sigma\  \ab_\nu(z') \epsilon^{\nu\lambda\beta}\del'_\lambda
[\fr{1}{\lambda^2} D^{\mu\nu}(z-z')] \}
\eqno(5.2)
$$
In the last term, the expression between brackets is the euclidean
propagator for the field $\ca_\mu$, where $D^{\mu\nu}$ is given by
(4.3). Inserting (4.3) in (5.2) we immediately
 find that only the first term contributes. After a
little manipulation
and analytic continuation back to the Minkowski space, we conclude that the
effective lagrangian for the $\ab_\mu$ field obtained by integration
over $\ca_\mu$ in (5.1) is
$$
\cl_{eff}[\ab_\mu] =-\fr{1}{4} \bar F_{\mu\nu}\Box^{-1/2}
\bar F^{\mu\nu}
\eqno(5.3)
$$
This is precisely the first term of the lagrangian (4.2) of the
pseudo electromagnetic field which, as we saw, describes correctly
the real electromagnetic interaction within a 2+1 D formulation.
We therefore are going to rewrite the lagrangian (4.2) in the
following way
$$
\cl_{PEM}[\ab_\mu ,\ca_\mu] =-\fr{\lambda^2}{4}  \cf_{\mu\nu}
\Box^{-1/2}  \cf^{\mu\nu} -(\fr{e}{2}j^\mu -\lambda \epsilon
^{\mu\alpha\beta}\del_\alpha\ca_\beta) \ab_\mu +\cl_{GF}
\eqno(5.4)
$$
As showed above, integration over $\ca_\mu$ will produce the lagrangian
(4.2). Instead of doing so, however, let us integrate over the
$\ab_\mu$ field, in order to get the effective lagrangian for
$\ca_\mu$. Adding the matter field kinetic lagrangian to (5.4) we have the
following vacuum functional (in Minkowski space)
$$
Z_{2+1} =Z_0^{-1} \int D\ab_\mu D\ca_\mu D\psi \exp \{i \int d^3z
[-\fr{\lambda^2}{4}\fr{\cf_{\mu\nu}^2}{\Box^{1/2}}
- (\fr{e}{2} j^\mu -\lambda \epsilon^{\mu\alpha\beta}\del_\alpha
\ca_\beta) \ab_\mu + \cl_{M} +\cl_{GF} ] \}
\eqno(5.5)
$$
where $\psi$ represents the matter fields.

\spa Integrating over the field $\ab_\mu$ in (5.5), as we promised,
we see that we produce a functional delta function identifying
the matter current with the topological current of the
$\ca_\mu$ field, namely
$$
Z_{2+1} =Z_0^{-1} \int D\ca_\mu D\psi\  \delta[\fr{e}{2} j^\mu -\lambda
\epsilon^{\mu\alpha\beta}\del_\alpha \ca_\beta ]
\exp \{i\int d^3z [-\fr{\lambda^2}{4}\fr{\cf^2_{\mu\nu}}{\Box^{1/2}}
+\cl_M +\cl_{GF} ] \}
\eqno(5.6)
$$
We see that the constraint generated in the theory which represents the
electromagnetic interaction of QED in a three-dimensional
formulation is precisely the Chern-Simons constraint (2.2) relating
the matter current to the statistical field. From (2.2) we
see that the relation between $\lambda$ and the Chern-Simons
parameter $\theta$ must be
$\theta =2\lambda$. This explains why we called $\ca_\mu$ the
``statistical field''.

\spa Observe that the only interaction between the matter field
and the $\ca_\mu$-field is the one induced by the constraint.
There is in particular no minimal coupling between these fields.
One can see directly from (5.6) that this peculiar interaction does
produce the correct electromagnetic interaction for the carged
matter. Indeed, writing the effective lagrangian for the
$\ca_\mu$-matter system as (with a slight abuse of notation
for the constraint)
$$
\cl[\ca_\mu, \psi] = \cl_M
-\fr{\lambda^2}{4}\fr{\cf^2_{\mu\nu}}{\Box^{1/2}} + constraint
[j^\mu \equiv \fr{2\lambda}{e} \epsilon^{\mu\alpha\beta}\del_\alpha\ca_\beta]
$$
$$
= \cl_M + \fr{\lambda^2}{2} \epsilon^{\mu\alpha\beta}\del_\alpha
\ca_\beta [\Box^{-1/2}] \epsilon_{\mu\sigma\lambda}\del^\sigma
\ca^\lambda + constraint[j^\mu\equiv \fr{2\lambda}{e}
\epsilon^{\mu\alpha\beta}\del_\alpha\ca_\beta]
\eqno(5.7)
$$
it is easy to see that
 integration over $\ca_\mu$ with the help of the delta function
constraint, will produce the following
 effective interaction for the matter field
$$
\cl_{eff}^{EM}=\cl_M - \fr{e^2}{8}j^\mu \Box^{-1/2} j_\mu
\eqno(5.8)
$$
 This, according to (4.1b), describes the correct electromagnetic
interaction for particles moving on a plane.

\spa Starting from QED in 3+1 D and restricting the
motion of the charged particles to a plane we have arrived at an effective
theory in 2+1 D which possesses the same constraint as the
Chern-Simons theory. We have seen in Section 2 that the statistics
transmutation taking place in Chern-Simons theory is a direct
consequence of this constraint. One is therefore naturally led to
ask whether the theory described by (5.7) should also induce
a statistical transmutation on the charged matter particles
coupled to $\ca_\mu$. The answer for this question is no. The
reason is that there is no minimal coupling between the matter
current $j^\mu$ and the $\ca_\mu$-field in (5.7). In order to
have statistical transmutation we need both the constraint (2.2)
and a minimal coupling to the matter current.

\spa  A simple way to prove the previous statement
 is to consider a static point charge minimally coupled to the
statistical field. Using the constraint, eq. (2.4), we see that there is
going to be a point statistical magnetic flux associated to
this static point charge. This configuration of the statistical magnetic
field corresponds to a vector potential $\ca_i =\del_i
arg(\vec x)$. If there is a minimal coupling between the current and the
vector potential $\ca_\mu$ we immediately realize that we can
eliminate this configuration by performing a gauge transformation
on the matter fields. As a consequence, these fields will acquire a
phase proportional to $arg(\vec x)$. It is clear that the
transformed fields will get an extra phase under rotations of
$2\pi$ after the transformation. In other words it will have
its spin/statistics changed. In the absence of the minimal coupling,
as in (5.7), on the other hand, the presence of the point magnetic
flux would not imply a change in statistics through the above
gauge transformation.

\spa Another way of seeing that the lagrangian (5.7) will not
produce a change in the statistics of the matter fields is to
 observe that the effective interaction in (5.8) does not contain
a term like the one appearing in (2.6) which is
the effective interaction responsible for the change in statistics.

\spa In the next section we are going to see how an explicit
Chern-Simons term for the $\ca_\mu$-field, as well as
a $j^\mu-\ca_\mu$ coupling, will be induced starting from
3+1 D.

\bigskip
\leftline{\Large\bf 6) Induction of the Chern-Simons Term}
\vskip 3pt
\leftline{\Large\bf and Statistical Transmutation}
\bigskip

\spa  Let us start by considering the topological $\theta$-term action
for the electromagnetic field, namely
$$
S_\theta =-\fr{\theta}{4} \int d^4x F_{\mu\nu}{\tilde F}^{\mu\nu}
\eqno (6.1)
$$
where $\theta$ is an arbitrary real number. This action is usually
uninteresting both at classical and quantum level. It is
the integral of a total
derivative
$$
S_\theta =-\fr{\theta}{2} \int d^4x\  \del_\mu I^\mu\ \ \ \ \ ;\ \ I^\mu =
\epsilon^{\mu\nu\alpha\beta}A_\nu\del_\alpha A_\beta
\eqno(6.2)
$$
and therefore it does not affect the equations of motion and bears
no effect on the classical behavior of the system.
At the quantum level, on the
other hand, the introduction of (6.1) in the fuctional
integral would in principle have consequences on the quantum behavior
of the system. It happens however that $S_\theta$ is a topological
invariant which classifies the homotopy classes of the $\Pi_3$ mappings
and for an abelian (U(1)) field like $A_\mu$ this mapping is always trivial.
 (This of course is not the case for nonabelian fields where the
$\theta$ term is known to lead to the nontrivial vacuum structure
of the theory \cite{tetavac}).
 We are going to see, however, that when we constrain the charged
particles to move on an infinite plane substrate as we have been
doing in the previous sections the $S_\theta$ term (6.1) does
produce a nontrivial effect on the dynamics of the effective
2+1 dimensional system. Indeed, considering the geometry depicted
in Fig.1 which is appropriate for the system we have been
  investigating and applying the Gauss theorem to (6.2), we get
$$
S_\theta =-\fr{\theta}{2}[ \int _{z_3=0}d^3\xi^\mu I_\mu +
 \int_{S_\infty} d^3\xi^\mu I_\mu ]
\eqno(6.3)
$$
Neglecting the term involving the surface at infinity, noting that
for the first term in (6.3) ( for the surface at $z^3 =0$)
$d^3\xi^\mu = -d^3\xi^3 = -dx^0dx^1dx^2$ and using the conventions
$\epsilon^{0123}=\epsilon_{3012}=1$ and $\epsilon^{012}=\epsilon
_{012}=1$
we immediately see that we obtain the Chern-Simons action
$$
S_\theta =\frac{\theta}{2} \int d^3x \epsilon^{\mu\alpha\beta}
A_\mu\del_\alpha A_\beta
\eqno(6.4)
$$
This fact was exploited in \cite{bau} where a four dimensional
representation of the Chern-Simons action was used.

 \spa We see that starting in 3+1 D from the action
$$
S_\theta + S_I = \int d^4x\ [-\fr{\theta}{4}F^{\mu\nu}{\tilde
F}_{\mu\nu} - e\ j^\mu_{3+1} A_\mu ]
\eqno(6.5)
$$
we arrive in 2+1 D at
$$
S = \int d^3x [\fr{\theta}{2} \epsilon^{\mu\alpha\beta}A_\mu
\del_\alpha A_\beta - e\ j^\mu A_\mu]
\eqno(6.6)
$$
This will induce the effective interaction which we found in (2.7)
and which is responsible for the statistical transmutation.

\spa  Going back to the lagrangian (4.2) which was obtained from
QED by dimensional reduction we see,
on the basis of the above reasoning, that when we  constrain
the system to the geometry of Fig. 1, we must also take into account
the $\theta$-term for the electromagnetic field. We therefore arrive at the
following effective lagrangian in 2+1 D
$$
\cl_{PEM}[\ab_\mu,A_\mu] =-\fr{1}{4}\bar F_{\mu\nu}(\Box^{-1/2})
\bar F^{\mu\nu} -\fr{e}{2}j^\mu\ab_\mu -e j^\mu A_\mu
+\fr{\theta}{2} \epsilon^{\mu\alpha\beta}A_\mu\del_\alpha A_\beta
\eqno(6.7)
$$
In this equation $A_\mu$ is the $\theta$-induced 2+1 D relic of the
electromagnetic field.

\spa Going to the $\ca_\mu$-field language  and integrating over
 $\ab_\mu$ as we did in section 5, we will immediately
obtain
$$
\cl[\ca_\mu,A_\mu]= -\fr{\lambda^2}{4} \fr{\cf^2_{\mu\nu}}{\Box^{1/2}}
+\cl_M + \fr{\theta}{2} \epsilon^{\mu\alpha\beta}
A_\mu\del_\alpha A_\beta - ej^\mu A_\mu+constraint[j^\mu\equiv \fr{2\lambda}{e}
\epsilon^{\mu\alpha\beta}\del_\alpha\ca_\beta]
\eqno(6.8)
$$
Substituting $j^\mu$ for the constraint in (6.8),
going to euclidean space and integrating over
$A_\mu$ with the help of the euclidean propagator (2.6) we generate
the following $\theta$-dependent term for $\ca_\mu$
$$
Z_{\theta}[\ca_\mu]=Z_0^{-1}\int DA_\mu \exp \{-
 \int d^3z_E [-\fr{i\theta}{2}\cs + 2i\lambda \epsilon^{\mu\alpha\beta}
A_\mu\del_\alpha \ca_\beta]
$$
$$
=\exp \{-2\lambda^2 \int d^3z\  d^3z'\ \ca_\mu(z) \epsilon^{\mu\sigma\alpha}
\del_\sigma \ca_\nu(z')\epsilon^{\nu\lambda\beta}\del_\lambda'
[\fr{i}{\theta}\epsilon^{\alpha\rho\beta}\del_\rho[\fr{1}{-\Box_E}]]\}
$$
$$
=\exp\{-i\fr{2\lambda^2}{\theta} \int d^3z_E\  \csi\}
\eqno(6.9)
$$
This expression is telling us that the $\theta$-term generates
a Chern-Simons term for the statistical field $\ca_\mu$ as well. Going to
Minkowski space and  adding the result of the integration
(6.9) to (6.8), we get the complete $\ca_\mu $ lagrangian
$$
\cl[\ca_\mu]=-\fr{\lambda^2}{4}\fr{\cf^2_{\mu\nu}}{\Box^{1/2}}
-\fr{2\lambda^2}{\theta}\csi +\cl_M+constraint[j^\mu\equiv
\fr{2\lambda}{e}\epsilon^{\mu\alpha\beta}\del_\alpha\beta]
\eqno(6.10)
$$
Making the choice $\lambda=\fr{\theta}{2}$, as we did in section 5,
 and using the constraint we may finally write
$$
\cl[\ca_\mu]=-\fr{1}{4}\cf_{\mu\nu}[\fr{\theta^2}{4\Box^{1/2}}]\cf^{\mu\nu}
+ \fr{\theta}{2}\csi -ej^\mu\ca_\mu +\cl_M
+constraint[j^\mu \equiv \fr{\theta}{e} \epsilon^{\mu\alpha\beta}
\del_\alpha\ca_\beta]
\eqno(6.11)
$$
This is our final expression for the complete Chern-Simons theory
including the true elctromagnetic interaction between charged
particles constrained to move on an infinite plane.
Using the constraint on the first term in (6.11) will immediately
produce the electromagnetic interaction (4.1b) or (3.5). Insertion
of the constraint in the second and third terms of (6.11),
on the other hand, and using the corresponding solution for $\ca_\mu$
in terms of $j^\mu$, Eq. (2.3), will produce the statistical
interaction which we have in (2.7). Indeed, integration over
the $\ca_\mu$-field with the help of the delta function constraint yields
the following effective lagrangian for the charged matter
$$
\cl_M^{eff}=\cl_M - \fr{e^2}{8}j^\mu\fr{1}{\Box^{1/2}}j_\mu
+ \fr{e^2}{2\theta} \epsilon^{\mu\alpha\beta}j_\mu\del_\alpha
[\fr{1}{\Box}]j_\beta
\eqno(6.12)
$$
The first term is the kinetic lagrangian and is completely
general.The second term is the electromagnetic interaction and the third
term is the satatistical interaction induced by the $\theta$-term
in the 2+1 D space.

\spa The presence of the minimal coupling  along with
the constraint in (6.11) indicate that by the gauge transformation described at
the end of section 5 the charged fields will suffer a change
$\Delta S=\fr{e^2}{2\pi\theta}$ in their spin/statistics.

\spa
Apart from reproducing the true electromagnetic interaction within
the framework of 2+1 dimensional
space-time (and in particular the 1/r Coulomb potential between
static point carges) the theory described by (6.11) has a number
of very interesting features.  The $\ca_\mu$-field part of the
lagrangian (6.11) is precisely what one obtains as
the bosonic field lagrangian in the bosonization
of the three-dimensional Dirac fermion field in 2+1 D \cite{bos}.
It is quite suggestive that it appears here in a different
context but also associated with a change in statistics.
The full consequences of this connection where certainly
not yet completely explored and would be worthwile to
understand more profoundly.

\spa  The quantization of theories possesing the nonlocality
of the type appearing in (6.11) was studied in detail in
\cite{rub}. A nice  property of (6.11) is that the
propagator of the $\ca_\mu$-field (for retarded or advanced
prescriptions) has support on the light-cone surface as in the
case of photons in 3+1 D and therefore the theory obeys the
 Huygens principle in the same way as four-dimensional QED
but unlike its three-dimensional counterpart \cite{rub,gia}.
 This observation allows us to conclude that in spite
of the nonlocality of the first term in (6.11) the theory
does respect causality, because by choosing retarded or advanced
prescriptions, for instance, one can show that the kernel
$\Box^{-1/2}$, the same appearing in the $\ca_\mu$ propagator,
 has support on the light cone surface \cite{rub,gia}.

\vfill
\eject

\leftline{\Large\bf 7) Conclusions}
\bigskip

\spa We have seen that the electrodynamics of particles moving
on an infinite plane leads to an effective 2+1 dimensional theory
which possesses the Chern-Simons constraint relating the
matter current to the gauge field. The geometry of the system,
on the other hand,
implies that the contribution of the topological $\theta$-term
can be no longer neglected. An explicit Chern-Simons term for
the 2+1 dimensional gauge field as well as a coupling of the
matter to it are then induced  in the effective 2+1 dimensional
theory. It is important to remark that the Chern-Simons
induction we consider here is completely different from
the one which is produced by integration over the matter
fields \cite{csgeneration}. The Chern-Simons term we found
here coexists with the matter fields.

 \spa The effective theory we obtained
completely describes the real electromagnetic
interaction of the charged particles, in spite of being three
dimensional and provides therefore a very convenient framework
for the description of realistic Condensed Matter systems like
the electron gas undergoing the Quantum Hall Effect or
maybe the High-Tc Superconductors.
It would be extremely interesting to investigate the behavior
of the theory introduced here in the presence of an external
electromagnetic field. In the case of a constant external
magnetic field, for instance, one would have the situation relevant for
the Quantum Hall Effect. It is probable that the magnetic field
would tune the statistical parameter $\theta$ in such a way
as to produce the observed effects in the conductance, for
instance. We are presently investigating this question.

\vskip 10mm

\leftline{\Large\bf Acknowledgments}
\bigskip

I would like to thank the Physics Department of Princeton University
and especially C.Callan and D.Gross for the kind hospitality. This work
was
supported in part by the Brazilian National Research Council (CNPq).

\vfill
\eject

\centerline{\Large\bf Figure Caption}

\vskip 2.0cm

\spa Fig.1 \spa Geometry of the system  of charged particles moving on
an infinite plane and the surface used in the Gauss theorem (dashed
line).

\end{document}